\documentclass[aps,prl,twocolumn,floatfix,superscriptaddress,showpacs]{revtex4}
\usepackage{graphicx}
\usepackage[pdftex,colorlinks,citecolor=blue,bookmarks]{hyperref}
\usepackage{bm}

\begin{document}
\title{Energy density functional study of nuclear matrix elements
  for neutrinoless $\beta\beta$ decay}

\author{Tom\'as R. Rodr\'iguez} 
\affiliation{GSI Helmholtzzentrum
  f\"ur Schwerionenforschung, D-64259 Darmstadt, Germany}
\affiliation{Departamento de F\'isica Te\'orica, Universidad
  Aut\'onoma de Madrid, E-28049 Madrid, Spain} 
\affiliation{CEA, Irfu, SPhN, Centre de Saclay, F-911191
  Gif-sur-Yvette, France} 
\author{Gabriel Mart\'{i}nez-Pinedo} 
\affiliation{GSI Helmholtzzentrum f\"ur
  Schwerionenforschung, D-64259 Darmstadt, Germany}
\date{\today} 
\pacs{21.60.Jz, 23.40.-s, 23.40.Hc}
\begin{abstract}
  We present an extensive study of nuclear matrix elements (NME) for
  the neutrinoless double beta decay of the nuclei $^{48}$Ca,
  $^{76}$Ge, $^{82}$Se, $^{96}$Zr, $^{100}$Mo, $^{116}$Cd, $^{124}$Sn, $^{128}$Te,
  $^{130}$Te, $^{136}$Xe, and $^{150}$Nd based on state-of-the-art
  energy density functional methods using the Gogny D1S
  functional. Beyond mean-field effects are included within the
  generating coordinate method with particle number and angular
  momentum projection for both initial and final ground states. We
  obtain a rather constant value for the NME's around 4.7 with the
  exception of $^{48}$Ca and $^{150}$Nd, where smaller values are
  found. We analyze the role of deformation and pairing in the
  evaluation of the NME and present detailed results for the decay of
  $^{150}$Nd.
\end{abstract}
\maketitle

Double beta decay is an extremely rare process where an even-even
nucleus decays into the even-even neighbor bypassing the energetically
forbidden odd-odd intermediate isobar. Along the nuclear chart, there
are only few candidates, all found in the valley of
stability~\cite{Avignone.Elliott.Engel:2008}. The double beta decay
where two electrons and two neutrinos are emitted in the final state
($2\nu\beta\beta$) has been observed experimentally in several
isotopes with half-lives $\sim10^{19-21}$~yr. This process conserves
the leptonic number and is compatible with Majorana or Dirac
neutrinos. There is also a second mode without neutrino emission
($0\nu\beta\beta$) that is possible only if the neutrinos are massive
Majorana particles and is related to the absolute mass scale of these
elementary particles~\cite{Avignone.Elliott.Engel:2008}. Except to one
controversial claim \cite{Klapdor-Kleingrothaus.Krivosheina.ea:2004},
$0\nu\beta\beta$ decay has not been detected and is currently the main
goal of several projects worldwide \cite{Ejiri:2010}. The half-life of
this process between $0^{+}$ states for mother and granddaughter
nuclei can be written as \cite{Avignone.Elliott.Engel:2008}:

\begin{equation}
  \label{halflife}
  \left[T^{0\nu}_{1/2}(0^{+}\rightarrow 0^{+})\right]^{-1}=G_{01}\left|M^{0\nu}\right|^{2}\left(\frac{\langle m_{\beta\beta}\rangle}{m_{e}}\right)^{2}
\end{equation}
where $\langle m_{\beta\beta}\rangle$ is the effective Majorana
neutrino mass, $m_{e}$ is the electron mass, $G_{01}$ is a kinematical
phase space factor, and finally, $M^{0\nu}$ is the nuclear matrix
element (NME). Eq.~\ref{halflife} shows that a precise determination
of the effective neutrino mass requires, apart from the experimental
measurement of the $0\nu\beta\beta$ half-life, a reliable calculation
of the NME. So far, several nuclear structure methods have been
employed. The most used among them are the Interacting Shell Model
(ISM)~\cite{Caurier.Menendez.ea:2008,Menendez.Poves.ea:2009},
proton-neutron Quasi-Random Phase Approximation
(QRPA)~\cite{Simkovic.Pantis.ea:1999,Simkovic.Faessler.ea:2008,Kortelainen.Suhonen:2007,Suhonen.Civitarese:2010},
and, more recently, Projected Hartree-Fock-Bogoliubov (PHFB) in limited
configuration spaces and using schematic Pairing plus Quadrupole
interactions~\cite{Chaturvedi.Chandra.ea:2008,Rath:2009} and Interacting Boson
Model (IBM)~\cite{Barea.Iachello:2009}.

In this letter, we present major improvements with respect to previous PHFB calculations~\cite{Chaturvedi.Chandra.ea:2008,Rath:2009}: we use state-of-the-art density functional methods based on the well established Gogny
D1S functional~\cite{Berger.Girod.Gogny:1984} and a much larger single particle basis (eleven major oscillator shells); we perform particle number and angular momentum projection for both mother and granddaughter nuclei; and include configuration mixing within the generating
coordinate method (GCM)
framework~\cite{Bender.Heenen.Reinhard:2003,Rodriguez.Egido:2007}. All these developments have been shown to be necessary for an unified description of nuclear structure~\cite{Bender.Heenen.Reinhard:2003}. In particular, particle number projection before variation is fundamental for a correct treatment of pairing correlations~\cite{Rodriguez.Egido:2007} that play a very important role in $0\nu\beta\beta$ decay~\cite{Caurier.Menendez.ea:2008}.



The $0\nu\beta\beta$ NME can be separated into three terms: Fermi (F),
Gamow-Teller (GT) and Tensor (T)~\cite{Avignone.Elliott.Engel:2008}:
\begin{equation}
  M^{0\nu}=-\left(\frac{g_{V}}{g_{A}}\right)^{2}M^{0\nu}_{F} +
  M^{0\nu}_{GT}-M^{0\nu}_{T}  
  \label{MMM}
\end{equation}
with $g_{V}=1$ and $g_{A}=1.25$. The tensor term is small according to
the ISM and QRPA
calculations~\cite{Menendez.Poves.ea:2009,Kortelainen.Suhonen:2007}
and will be neglected in this work. We use the closure
approximation~\cite{Avignone.Elliott.Engel:2008} to sum over
intermediate states in the odd-odd nucleus as currently it is not
possible to compute odd-odd nuclei using beyond mean-field methods:
symmetry restoration methods including blocking effects have not been
fully developed so far. 
Therefore, a calculation of the $2\nu\beta\beta$ mode cannot be performed as the closure approximation is not valid in this process.
The
different terms in Eq. \ref{MMM} can be expressed as the expectation
value of a two-body operator between the initial and final states,
i.e. $M^{0\nu}_{F/GT}=\langle
0^{+}_{f}|\hat{M}^{0\nu\beta\beta}_{F/GT}|0^{+}_{i}\rangle$. 
Detailed expressions for $\hat{M}^{0\nu\beta\beta}_{F/GT}$ can be
found in Ref. \cite{Menendez.Poves.ea:2009}. We have included high order
currents~\cite{Simkovic.Pantis.ea:1999}, nucleon finite size
corrections~\cite{Simkovic.Pantis.ea:1999} and radial short range correlations treated
within the Unitary Correlator Method~\cite{Feldmeier.Neff.ea:1998,Kortelainen.Suhonen:2007}.
In the GCM+PNAMP approach (shortly GCM from now on), the initial ($i$)
and final ($f$) many-body wave functions are found as linear
combinations of particle number $N,Z$ and angular momentum $I=0$
projected wave functions with different intrinsic quadrupole
deformations $\beta$:
\begin{equation}
  |0^{+}\rangle=\sum_{\beta}g_{\beta}P^{I=0}P^{N}P^{Z}|\Phi_{\beta}\rangle
\end{equation}
where $P^{I=0}$, $P^{N},P^{Z}$ are the corresponding angular momentum
($I=0$) and particle number projectors~\cite{Ring.Schuck:1980}. The
intrinsic axial symmetric Hartree-Fock-Bogoliubov (HFB) wave functions
$|\Phi_{\beta}\rangle$ are solutions to the variation after particle
number projection equations constrained to a given value of the axial
quadrupole deformation,
$\beta$~\cite{Anguiano.Egido.Robledo:2001,Rodriguez.Egido:2007}. Therefore,
intrinsic deformation of the system is naturally included in the
formalism and pairing correlations properly taken into
account. Finally, the coefficients $g_{\beta}$ are found by solving
the Hill-Wheeler-Griffin (HWG) equation
\cite{Ring.Schuck:1980,Rodriguez.Egido.Robledo:2002}. First, for each nucleus we transform the
non-orthogonal set of wave functions
$\left\{P^{I=0}P^{N}P^{Z}|\Phi_{\beta}\rangle\right\}$ into an
orthonormal one
$\left\{|\Lambda\rangle=\sum_{\beta}\frac{u_{\Lambda,\beta}}{\sqrt{n_{\Lambda}}}P^{I=0}P^{N}P^{Z}|\Phi_{\beta}\rangle\right\}$
by diagonalizing the norm overlap matrix,
$\sum_{\beta'}\langle\Phi_{\beta}|P^{I=0}P^{N}P^{Z}|\Phi_{\beta'}\rangle
u_{\Lambda,\beta'}=n_{\Lambda}u_{\Lambda,\beta}$.  In this basis, the
HWG equation reads: $\sum_{\Lambda'}\varepsilon_{\Lambda\Lambda'}
G^{a}_{\Lambda'}=E^{a}G^{a}_{\Lambda}$, where
$\varepsilon_{\Lambda\Lambda'}$ are the so-called energy
kernel~\cite{Rodriguez.Egido:2007,Rodriguez.Egido.Robledo:2002}.  Finally, the coefficients for the
lowest eigenvalue are used to compute both the so-called collective
wave functions $F(\beta)=\sum_{\Lambda}
G^{0}_{\Lambda}u_{\Lambda,\beta}$ - probability distribution for the
state to have a given deformation - and the NME:
\begin{widetext}
  \begin{equation}
    M^{0\nu}_{F/GT}=\sum_{\Lambda_{i}\Lambda_{f}}\sum_{\beta_{i}\beta_{f}}
    \left(\frac{u^{*}_{\Lambda_{f},\beta_{f}}}{\sqrt{n_{\Lambda_{f}}}}\right) 
    G^{0*}_{\Lambda_{f}} \langle \Phi_{\beta_{f}}| P^{N_{f}}
    P^{Z_{f}} \hat{M}^{0\nu}_{F/GT} P^{I=0}P^{N_{i}}P^{Z_{i}}|
    \Phi_{\beta_{i}} \rangle G^{0}_{\Lambda_{i}} 
    \left(\frac{u_{\Lambda_{i},\beta_{i}}}{\sqrt{n_{\Lambda_{i}}}}\right). 
    \label{NME_GCM}
  \end{equation}
\end{widetext}
Particle number conservation together with using large and identical configuration
spaces for protons and neutrons guarantees that
Ikeda's sum rule is fulfilled. Additionally, to our knowledge, this is
the first implementation of the GCM method for calculating transitions
between different nuclei including particle number symmetry
restoration in both the initial and final states.\\ \indent
\begin{figure}[b]
  \centering
  \includegraphics[width=\columnwidth]{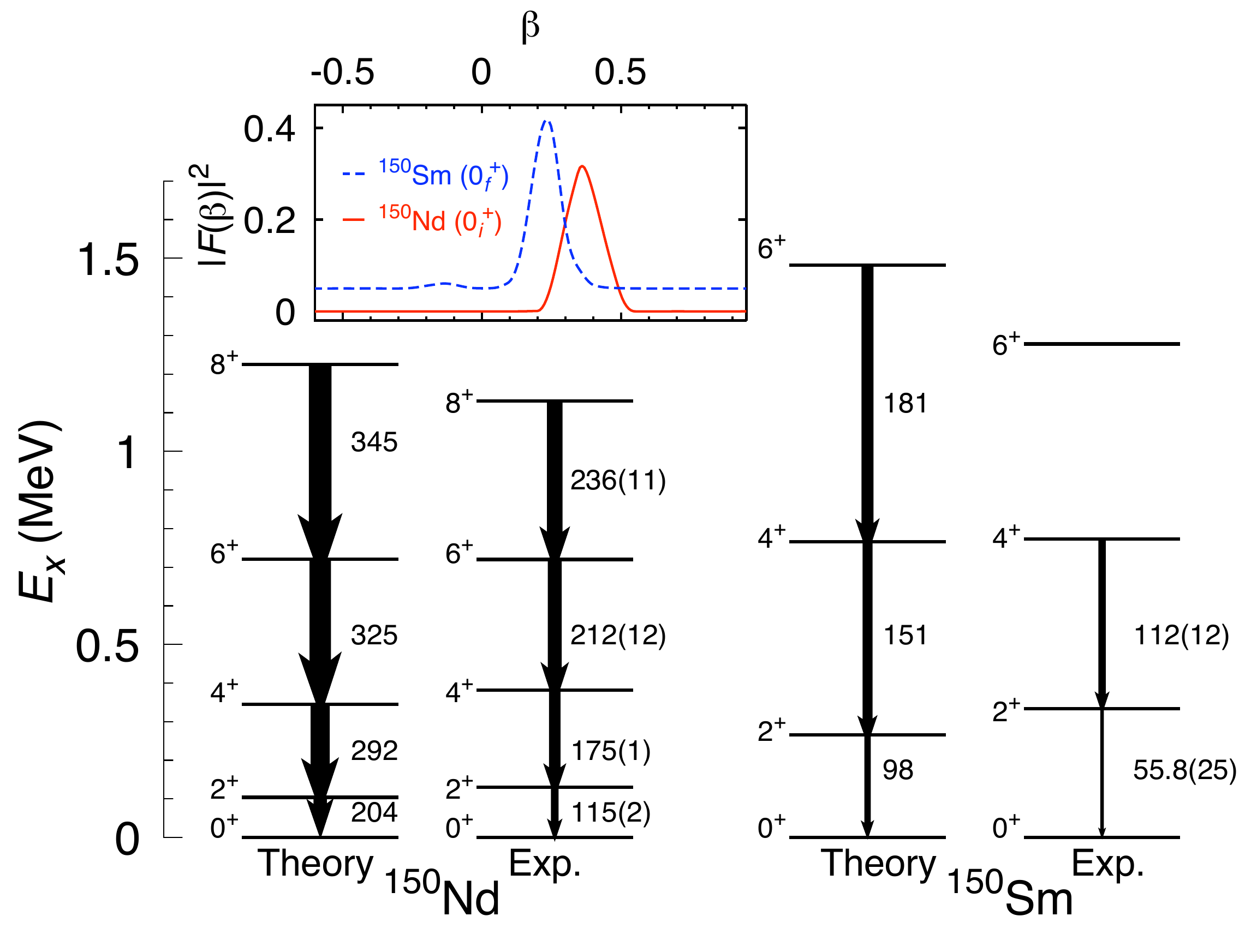}
  \caption{(color online) Comparison between theoretical and
    experimental ground state bands for $^{150}$Nd and
    $^{150}$Sm. Inset: collective wave functions of the ground states
    as a function of the intrinsic deformation. The values for
    $^{150}$Sm have been shifted up by 0.05.}
  \label{Fig1}
\end{figure}
We now present our results for the $0\nu\beta\beta$ NME discussing in
detail the decay of $^{150}$Nd. To check the reliability of the method
for describing properties of the initial and final nuclei, we show in
Fig.~\ref{Fig1} a comparison between the experimental and theoretical
ground state bands for $^{150}$Nd and $^{150}$Sm. We observe a rather
good agreement for both excitation energies and $B$(E2) transition
probabilities, with the theoretical results predicting a slightly
larger rotational (collective) character than the experiment. The
computed double beta decay $Q$-value is 2.99~MeV while the
experimental value is 3.37~MeV. The inset of Fig.~\ref{Fig1} shows the
probability distribution for the mother and granddaughter $0^{+}_{1}$
states to have a given intrinsic quadrupole deformation $\beta$. Both
$^{150}$Nd and $^{150}$Sm have well deformed prolate ground states
with $\beta\approx+0.40$ and $\beta\approx+0.25$, respectively. These
values are compatible with the rotational bands shown in the
figure. 
\begin{figure}[t]
  \centering

  \includegraphics[width=\columnwidth]{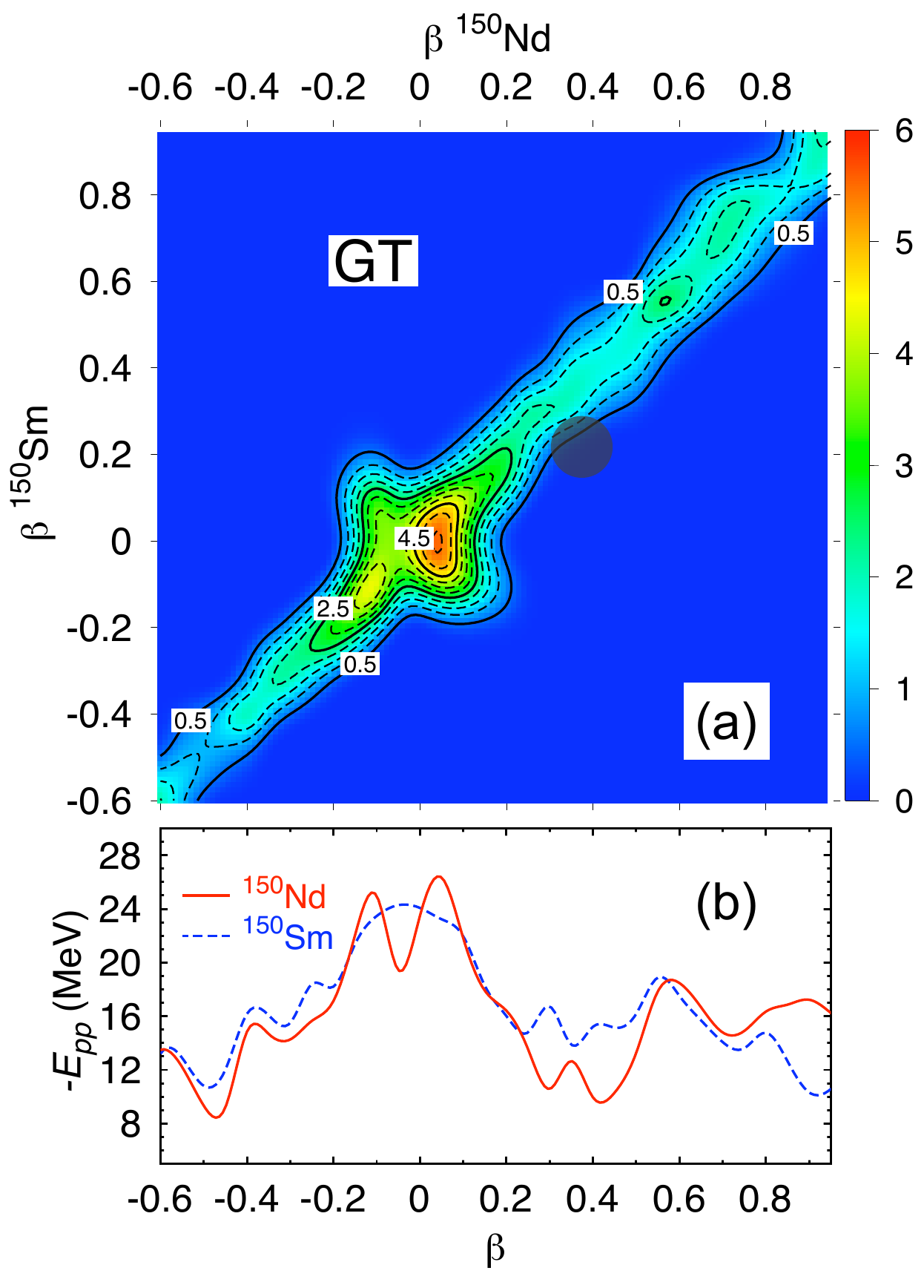}
  \caption{(Color online) (a) Strength of the GT operator as a
    function of the intrinsic deformation of the mother and
    granddaughter nuclei. Shaded area corresponds to the maximum probability
    for finding the mother and granddaughter states. (b) Pairing
    energies as a function of the intrinsic deformation.}
  \label{Fig3}
\end{figure}
Equation~\ref{NME_GCM}, shows that the $0\nu\beta\beta$ NME can be
expressed as a sum of matrix elements between states of different
intrinsic quadrupole deformation for the initial and final
nuclei. Consequently, the matrix element
$\frac{\langle\Phi_{\beta_{f}}|P^{N_{f}}P^{Z_{f}}\hat{M}^{0\nu\beta\beta}_{F/GT}
  P^{I=0}P^{N_{i}}P^{Z_{i}} | \Phi_{\beta_{i}} \rangle}{\sqrt{\langle
    \Phi_{\beta_{f}} |P^{N_{f}}P^{Z_{f}}P^{I=0}| \Phi_{\beta_{f}}
    \rangle \langle\Phi_{\beta_{i}}|P^{N_{i}}P^{Z_{i}}P^{I=0}|
    \Phi_{\beta_{i}} \rangle}}$ provides explicitly the strength of
the $0\nu\beta\beta$ operators as a function of the deformation of the
nuclei involved in the decay. In Fig.~\ref{Fig3}(a) we show these
matrix elements for GT component (the Fermi part gives a similar
pattern and it is not shown). The strength is concentrated rather
symmetrically in the diagonal part of the figure, implying that the
decay between states with different initial and final deformation is
hindered. Moreover, spherical configurations are the most preferred to
decay. Nevertheless, Fig. \ref{Fig3}(a) shows also an interval of
deformation close to the spherical configuration where non-negligible
off-diagonal matrix elements (greater that 0.5) are obtained
($\beta\in\left[-0.2,+0.2\right]$). The absolute maximum is found at
$(+0.03,0)$ with additional local maxima, for example at
$(+0.52,+0.52)$. These results are in agreement with ISM and PHFB
calculations that have shown a significant decrease of the NME with
increasing difference in quadrupole deformation between initial and
final states~\cite{Menendez.Poves.ea:2008,Chaturvedi.Chandra.ea:2008}. This trend has also been observed in
$2\nu\beta\beta$ QRPA calculations~\cite{Alvarez-Rodriguez.Sarriguren.ea:2004}. The
final value of the NME is determined by weighting the strength of the
transition operator with the wave functions of the initial and final
states which selects the relevant region of deformation. This area is
marked by a shaded circle in Fig. \ref{Fig3}(a) showing that in this
case both the difference in deformation and the absence of shape
mixing inhibit the transition. The final result for the NME is
$M^{0\nu}=1.71$ with 1.28 and 0.43 coming from Gamow-Teller and Fermi
components, respectively.  In order to shed additional light on the
structure of the strength we show in Fig. \ref{Fig3}(b) the pairing
energy $-E_{pp}$ of the nuclei involved in the decay. This energy is
defined as the pairing tensor part~\cite{Ring.Schuck:1980,Bender.Heenen.Reinhard:2003}
within the PNAMP approach. 
We find a strong correlation between the
structure of the NME and $-E_{pp}$, both having maxima at similar
values of the deformation for mother and granddaughter nuclei. This
result is also in agreement with ISM and QRPA calculations where the
largest values for the NME are obtained with well paired wave
functions with large zero seniority components in the spherical
basis~\cite{Caurier.Menendez.ea:2008,Simkovic.Faessler.ea:2008}.
\begin{figure}[t]
  \centering
  \includegraphics[width=\columnwidth]{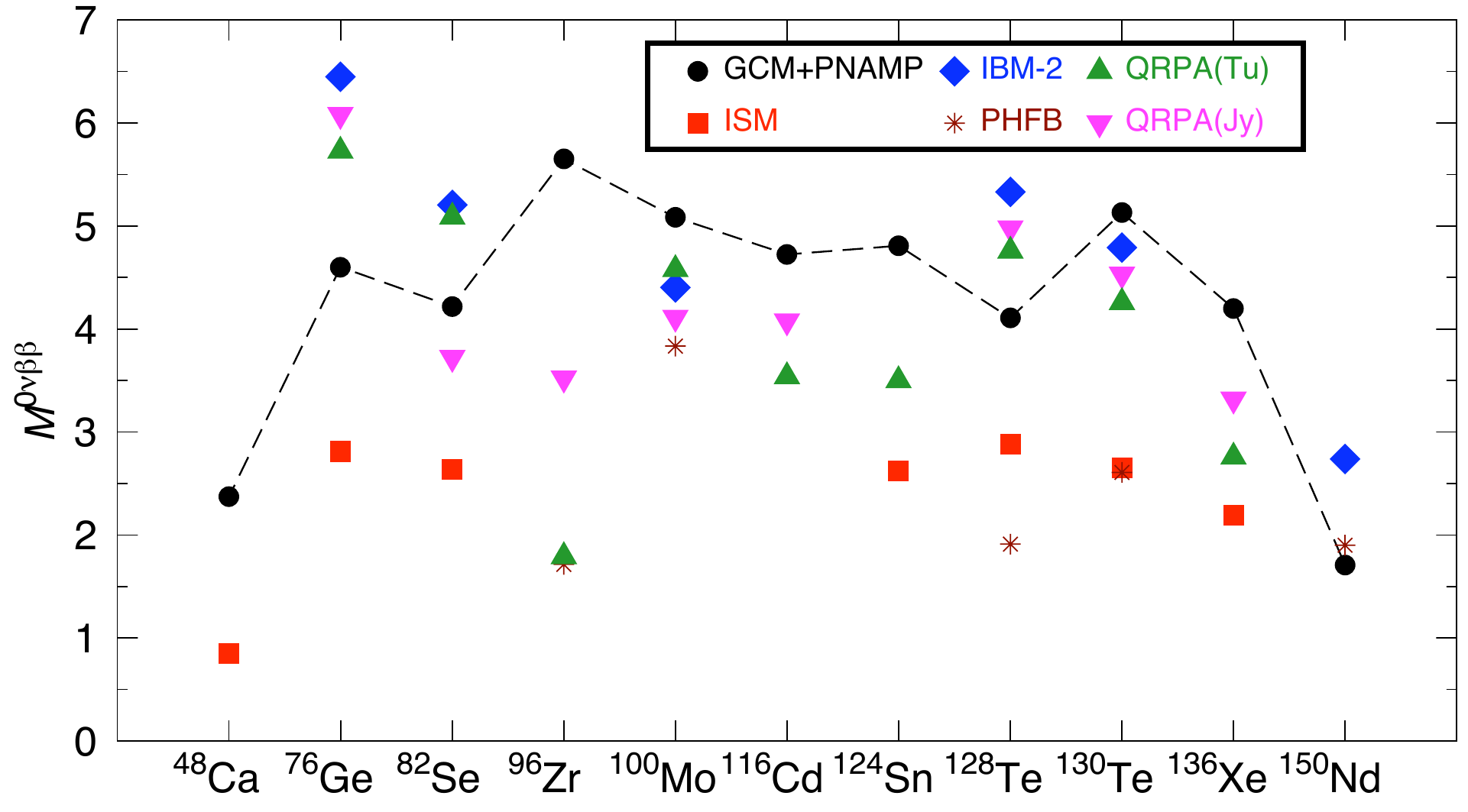}
  \caption{(Color online) Nuclear matrix elements calculated for
    different methods (ISM
    \cite{Menendez.Poves.ea:2009,Menendez.Poves.ea:2009b}, QRPA(Jy)
    \cite{Kortelainen.Suhonen:2007}, QRPA(Tu)
    \cite{Simkovic.Faessler.ea:2008}, IBM-2 \cite{Barea.Iachello:2009}, PHFB
    \cite{Chaturvedi.Chandra.ea:2008}) with UCOM short range
    correlations. QRPA values are calculated with $g_{A}=1.25$ and
    IBM-2 and PHFB results are multiplied by 1.18 to account for the
    difference between Jastrow and UCOM~\cite{Poves.priv}.} 
  \label{Fig4}
\end{figure}
\\ \indent We now present the values of the NME (Eq. \ref{NME_GCM})
computed for several double-beta emitters and compare with the ones
calculated with other methods that use similar assumptions concerning
the neutrino potentials. In Fig. \ref{Fig4} we observe that the NME's
are rather constant around an averaged value of 4.7 for the decay of
$^{76}$Ge, $^{82}$Se, $^{96}$Zr, $^{100}$Mo, $^{116}$Cd, $^{124}$Sn, $^{128}$Te,
$^{130}$Te and $^{136}$Xe, being $^{96}$Zr the one with the largest
value. In these nuclei, the differences in deformation between the
initial and final states are not very significant and also the
structure and absolute values of the transition strength are quite
similar. There are two exceptions to this general trend, namely,
$^{48}$Ca and $^{150}$Nd where the NME is significantly smaller. In
the later, the difference between deformation of the mother and
granddaughter is remarkable (inset Fig. \ref{Fig1}) and this is
precisely the main source of suppression of the NME. On the other
hand, due to the double magic character of $^{48}$Ca, this nucleus has
the smallest value for the pairing energy and transition strength
among the nuclei studied in this work. 
Concerning the NME's obtained by other methods, we attribute first the small NMEs obtained by PHFB to the
reduced configuration space used and the lack of particle number
restoration and shape mixing in those calculations. 
We also show that the
lowest NMEs are obtained within the ISM, a rather constant value
around 2.5 except for $^{48}$Ca. We expect that enlarging the model
space used in the ISM calculations will enhance the values of their
NMEs~\cite{Caurier.Menendez.ea:2008,Menendez.Poves.ea:2009b,Suhonen.Civitarese:2010}. Considering higher seniority components
in our method by including quasiparticle excitations in the intrinsic
wave functions may reduce slightly our values.  Some differences are
also found between the GCM and QRPA values. The main differences
between these approaches are the assumption of spherical symmetry in
QRPA, the absence of quasiparticle excitations in the GCM approach and
the much larger -and no core- single particle basis used in GCM. 
Furthermore, neither triaxial, mirror, nor time reversal symmetry
breaking effects are included in our GCM calculations because they are beyond current computational capabilities. We expect that the inclusion of these degrees of freedom will not change significantly the structure of the ground states as the nuclei studied here are either spherical or well deformed. To validate our approach we have computed the total GT strengths $S_{+(-)}$ for the decay of granddaughter (mother) nuclei defined, for example, in Ref. \cite{SM_RMP}. In Table~\ref{table}, we compare the calculated $S_{+(-)}$ with the experimental
values measured in charge exchange reactions. As in QRPA
and ISM calculations, a quenching factor of $(0.74)^{2}$ has been
introduced \cite{SM_RMP,Alvarez-Rodriguez.Sarriguren.ea:2004}. 
Finally, we evaluate the
half-life of each nuclei based on the NME calculated with GCM
method. In Table~\ref{table}, we show the difference between the
calculated and experimental $Q$-values and we observe an excellent
agreement in most of the cases except for $^{96}$Zr and $^{100}$Mo,
where an over binding of $^{96}$Mo and $^{100}$Ru isotopes gives such
differences. These are precisely the decays with largest NME and
smallest half-lives predicted by our calculations.

\begingroup
\squeezetable
\begin{table}[htb]
  \centering
  \begin{tabular}{cccccc}
    \hline
    \hline
    $A$ & $Q_{\text{theo}}-Q_{\text{exp}}$ &  $S_{+}$ & $S_{-}$ &
    $M^{0\nu}$ & $T_{1/2}$ \\
       &  (MeV) & & & & ($\times 10^{23}$~y)\\
    \hline
    48 & 0.265 & 1.99  & 13.55 & 2.37 & 28.5 \\
  & &  (1.9 $\pm$ 0.5 \cite{CaCER_09}) & (14.4$\pm$2.2 \cite{CaCER_09}) & & \\
    76 & 0.271 & 1.49  & 20.97 & 4.60 & 76.9 \\     
     & & (1.45 $\pm$ 0.07 \cite{SeCER_97}) & (19.89 \cite{Anderson_89}) & & \\     
    82 & $-$0.366 & 1.24 & 23.56  & 4.22 & 20.8 \\
    & & & (21.91 \cite{Anderson_89}) & & \\
    96 & 2.580 & 2.56   & 27.63 & 5.65 & 5.48  \\   
    & & (0.29 $\pm$ 0.08 \cite{MoCER_08}) & & &  \\   
    100 & 1.879 & 2.48 & 27.87 & 5.08 & 8.64 \\    
    & & & (26.69 \cite{Anderson_89}) & & \\    
    116 & 1.365 & 2.61 & 34.30 & 4.72 & 9.24 \\
    & & (1.09$^{+0.13}_{-0.57}$ \cite{SnCER_05}) & (32.70 \cite{Anderson_89}) & & \\
    124 & $-$0.830 & 1.63 & 40.65 & 4.81 & 16.2 \\       
    128 & $-$0.564 & 1.45 & 40.48 & 4.11 & 343.1 \\   
    & & & (40.08 \cite{Anderson_89}) & & \\   
    130 & $-$0.348 & 1.19 & 43.57& 5.13 & 8.84 \\     
    & & & (45.90 \cite{Anderson_89}) & & \\     
    136 & $-$1.027 & 0.96 & 46.71 & 4.20 &12.7 \\
    150 & $-$0.380 & 1.45 & 50.32 & 1.71 & 16.5 \\
    \hline
    \hline
  \end{tabular}
  \caption{Difference between theoretical and experimental $Q$ values,
    total GT strength for granddaughter (mother) $S_{+}$ ($S_{-}$) , NME and predicted half-lives for
    several $0\nu\beta\beta$ decaying nuclei assuming $\langle
    m_{\beta\beta}\rangle=0.5$ eV.} 
  \label{table}
\end{table}
\endgroup
In summary, we have presented the first calculations of
$0\nu\beta\beta$ decay within the energy density functional framework
including beyond mean-field effects. We have analyzed the role of the
intrinsic quadrupole deformation and pairing content of the nuclei
involved in this process. Decays between spherical initial and final
shapes are found to be favored while large differences in deformation
hinder significantly the transition probability. Our calculations
constitute the first consistent evaluation of the $0\nu\beta\beta$
decay of $^{150}$Nd. 
\begin{acknowledgments}
  We thank A. Poves, J. Men\'{e}ndez, J.L. Egido, K. Langanke,
  T. Duguet and F. Nowacki for fruitful discussions. TRR is supported by the Programa de Ayudas para Estancias de Movilidad
  Posdoctoral 2008 and FPA2009-13377-C02-01 (MICINN). GMP is partly supported by the DFG through contract SFB 634, by the ExtreMe
  Matter Institute EMMI and by the Helmholtz International Center for
  FAIR.
\end{acknowledgments}

\end{document}